\PassOptionsToPackage{table,xcdraw}{xcolor}
\documentclass[table,twocolumn]{aastex61}
\usepackage[table,xcdraw]{xcolor}
\usepackage{amstext}
\usepackage{amsmath}
\usepackage{amssymb}
\usepackage{apjfonts}
\usepackage{graphicx}
\usepackage{tikz}
\usepackage[capbesideposition={top}]{floatrow}
\usepackage{float}
\usetikzlibrary{decorations.pathreplacing,angles,quotes}

\newcommand{\beqar}{\begin{eqnarray}}
\newcommand{\eeqar}{\end{eqnarray}}

\newcommand{\beq}{\begin{equation}}
\newcommand{\eeq}{\end{equation}}

\definecolor{nick}{HTML}{006400}

\begin{document}
\title{Illuminating black hole subsystems in young star clusters}

\correspondingauthor{Nicholas Kaaz}
\email{nkaaz@u.northwestern.edu}

\author[0000-0002-5375-8232]{Nicholas Kaaz}
\affiliation{Department of Physics \& Astronomy, Northwestern University, Evanston, IL 60202, USA}
\affiliation{Center for Interdisciplinary Exploration \& Research in Astrophysics (CIERA), Evanston, IL 60202, USA}

\author[0000-0002-4086-3180]{Kyle Kremer}
\affiliation{Department of Physics \& Astronomy, Northwestern University, Evanston, IL 60202, USA}
\affiliation{Center for Interdisciplinary Exploration \& Research in Astrophysics (CIERA), Evanston, IL 60202, USA}

\author[0000-0002-4449-9152]{Katie Auchettl}
\affiliation{DARK, Niels Bohr Institute, University of Copenhagen,Blegdamsvej 17, DK-2100 Copenhagen, Denmark}
\affiliation{Department of Astronomy \& Astrophysics, University of California, Santa Cruz, CA 95064, USA}
\affiliation{School of Physics, The University of Melbourne, Parkville, VIC 3010, Australia}

\author[0000-0003-2558-3102]{Enrico Ramirez-Ruiz}
\affiliation{Department of Astronomy \& Astrophysics, University of California, Santa Cruz, CA 95064, USA}
\affiliation{DARK, Niels Bohr Institute, University of Copenhagen,Blegdamsvej 17, DK-2100 Copenhagen, Denmark}

\begin{abstract} 
    There is increasing evidence that globular clusters retain sizeable black hole populations at present day. This is supported by dynamical simulations of cluster evolution, which have unveiled the spatial distribution and mass spectrum of black holes in clusters across cosmic age. However, black hole populations of young, high metallicity clusters remain unconstrained. Black holes hosted by these clusters mass segregate early in their evolutionary history, forming central subsystems of hundreds to thousands of black holes. We argue that after supernova feedback has subsided ($\gtrsim 50\,{\rm Myr}$), the host cluster can accumulate gas from its dense surroundings, from which the black hole subsystem accretes at highly enhanced rates. The collective accretion luminosity can be substantial and provides a novel observational constraint for young massive clusters. We test this hypothesis by performing 3D hydrodynamic simulations where we embed discretized potentials, representing our black holes, within the potential of a massive cluster. This system moves supersonically with respect to a gaseous medium from which it accretes. We study the accretion of this black hole subsystem for different subsystem populations and determine the integrated accretion luminosity of the black hole subsystem. We apply our results to the young massive clusters of the Antennae Galaxies and find that a typical subsystem accretion luminosity should be in excess of $\approx 10^{40}\,{\rm ergs\,\,s^{-1}}$. We argue that no strong candidates of this luminous signal have been observed and constrain the subsystem population of a typical cluster in the Antennae Galaxies to $\lesssim10-2\times10^2$ $10\,M_\odot$ black holes, given that feedback doesn't significantly impede accretion and that the gas remains optically thin. 
\end{abstract}

\section{Introduction}
\label{sec:intro}

Understanding the populations of black holes in globular clusters (GCs) and other dense stellar systems is of considerable interest to the astrophysics community. This is because it provides insights on the stellar and dynamical evolution of clusters and informs constraints on the merger rates of binary black holes (BBHs). Classically, it was assumed that early in the evolutionary history of dense clusters, the black hole population would rapidly mass segregate, forming a dynamically unstable central subsystem. The dynamical interactions operating within this subsystem would lead to efficient ejection of the black hole members on sub-Gyr timescales, suggesting old systems such as GCs should retain only a small number of their black holes by present day \citep{spitzer_1987,kulkarni_1993,sigurdsson_hernquist_1993}. However, over the past decade there has been new evidence that has cast doubt on this assumption. In particular, there have been numerous detections of black hole candidates as members of binaries with luminous stellar companions in both extragalactic and galactic GCs. These binary candidates that have been identified through both X-ray and radio observations of accreting systems \citep{Maccarone2007,Strader2012,Shishkovsky2018} and, more recently, through dynamical measurements \citep{Giesers2018,Giesers2019}.

The retention of black holes in GCs has been further motivated by recent computational simulations of GC evolution which show that black holes play an essential role in the long-term dynamical evolution of clusters, particularly the process of cluster core collapse \citep[e.g.,][]{merritt_2004, mackey_2007, mackey_2008, breenheggie_2013a, breenheggie_2013b,morscher_2015, peuten_2016, wang_2016, arcasedda_2018, kremer_2018a, kremer_2019, kremer_2019c}. While a large number of BHs are retained in a GC, the cluster exhibits a large observed core radius due to energy generated through ``black hole burning'' \citep[the cumulative effect of black hole binary formation, hardening, and dynamical ejections; for a recent review see,][]{kremer_2019d}. Only when a GC's black hole population has been almost fully depleted can a cluster attain a core-collapsed architecture. Along these lines, several groups have suggested that cluster black hole populations (or lack thereof) can be indirectly inferred from structural features, such as a large core radius and low central density \citep[e.g.,][]{merritt_2004,Chatterjee_2017,Askar_2018, arcasedda_2018, kremer_2019}. Similarly, \citet{Weatherford_2018, Weatherford_2019} demonstrated that mass-segregation measurements can be used as a robust indicator of black hole populations in GCs.

Recent analyses have also proposed additional observations that may constrain populations of black holes in GCs. 
For example, tidal disruption events in GCs, which may be detectable by transient surveys such as ZTF and LSST may constrain black hole populations in dense cluster environments \citep{perets_2016,lopez_2019,samsing_2019,kremer_2019b}. It is also now understood that the dynamical processes operating in dense systems like GCs make them ideal factories for the formation of BBH mergers that may be detected as gravitational-wave (GW) sources by both current (LIGO/Virgo) and future (e.g., LISA) GW detectors \citep[e.g.,][]{Samsing2014,Rodriguez2015a,Rodriguez2016a,Askar2017,Samsing2017, Banerjee2017,Hong2018,Fragione2018b,Samsing2018a,Rodriguez2018b,Zevin2018,kremer_2019e,kremer_2019c}. Thus, GW observations may provide an additional constraint upon black hole populations in these systems.

While the question of black hole retention has been studied extensively in the context of late time ($\gtrsim10\,\,{\rm Gyr}$) environments, the fate of black hole populations in young environments ($<1\,\,{\rm Gyr}$), such as young massive clusters (YMCs) \citep{zwart_2010}, remains unclear. YMCs are often considered young GC analogues due to their similar masses ($\approx10^5-10^6\,M_\odot$) and core radii ($\approx0.5-1\,{\rm pc}$), but they are distinguished by having much higher metallicities ($Z\approx Z_\odot$). This suggests that their stellar populations lose mass from stellar winds more efficiently, truncating the resulting black hole mass spectrum. As a consequence, black holes in YMCs receive larger natal kicks, making it more difficult for the host cluster to retain them. In dynamical simulations of solar metallicity clusters, which use prescriptions for stellar wind mass loss and black hole natal kicks, it is typical that at ages of $\approx 1\,{\rm Gyr}$, $10^5-10^6\,M_\odot$ clusters retain populations of hundreds to thousands of black holes \citep{kremer_2019c}. However, because our understanding of black hole natal kicks and stellar winds is currently limited, our prescriptions for these processes are poorly constrained. This uncertainty makes it unclear if solar metallicity clusters can retain as many black holes at birth as recent studies might suggest, motivating us to search for new observational constraints. 
%\sout{YMCs, which have typical masses of $\approx10^5-10^6\,M_\odot$ and ages of $\approx1-500\,{\rm Myr}$, are often considered young GC analogues. However, they have the important distinction of having much higher metallicities ($Z\approx Z_\odot$) than GCs, suggesting that their stellar populations lose mass from stellar winds more efficiently, truncating the resulting black hole mass spectrum and causing the black holes to receive higher natal kick velocities. 

%When dynamical simulations of solar metallicity clusters prescribe stellar wind mass loss rates and black hole natal kick distributions, they find that at ages of $\approx 1\,{\rm Gyr}$ clusters at masses of $\approx 10^5-10^6\,M_\odot$ retain populations of hundreds to thousands of black holes \citep{kremer_2019c}. However, our current understanding of black hole natal kicks and stellar winds is poorly constrained, making it unclear if solar metallicity clusters can retain as many black holes at birth as these simulations suggest. This uncertainity limits our ability to predict black hole populations in YMCs theoretically, motivating us to instead search for new observational constraints. }

\begin{figure}[bht]
    \includegraphics[width=\textwidth]{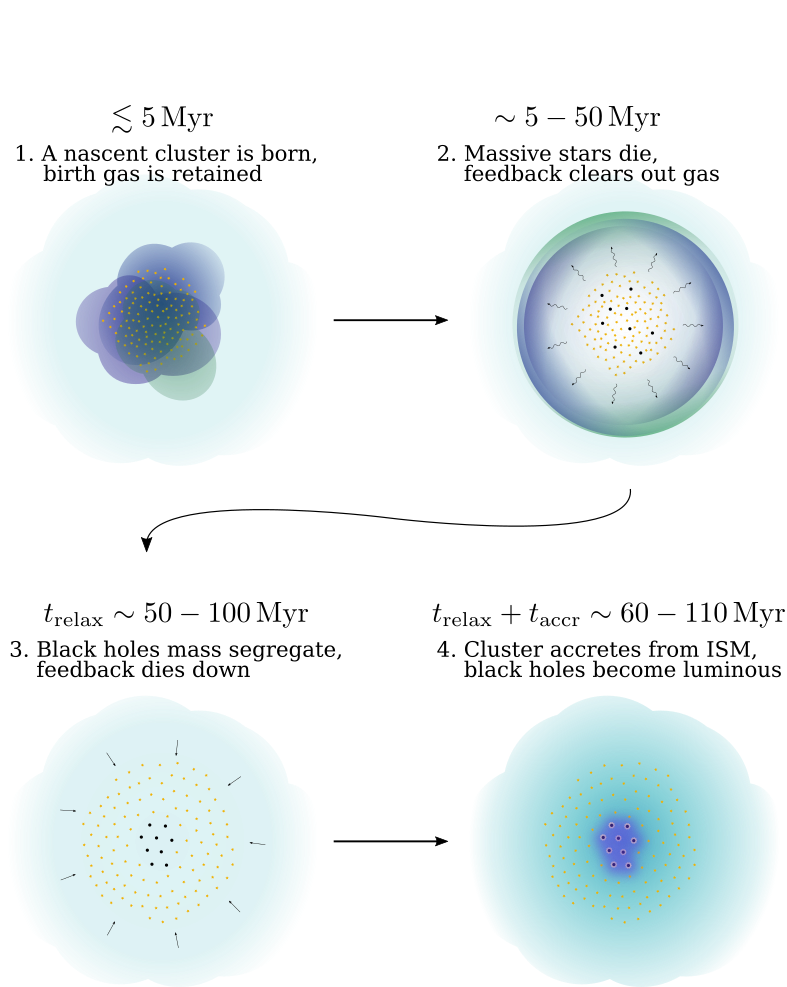}
    \caption{Here, we sketch the envisioned timeline of cluster gas retention concurrent with black hole mass segregation. \textit{1.} In the first few million years of the clusters lifetime, feedback processes are not yet significant, and the cluster retains its natal gas.  \textit{2.} Over the next $\approx 50\,{\rm Myr}$, massive stars evolve and die, supernova and stellar wind feedback strips the cluster of its natal gas, and the black hole population is born. \textit{3.} After roughly $\approx 50\,{\rm Myr}$, the majority of massive stars have evolved and died, feedback has died down, and the cluster can begin accumulating gas from the ISM. The black hole population mass segregates into a centrally-confined subsystem on the order of a relaxation timescale. \textit{4.} The black hole subsystem collectively accrete from the cluster gas accumulation, producing a luminous signal. Black holes will be gradually ejected from the subsystem, and by $\approx 10$ relaxation times ($\approx 1\,{\rm Gyr}$), the subsystem will no longer exist, and the proposed luminous signal will no longer be viable.}
    \label{fig:cartoon}
\end{figure}

Let us consider a dense star cluster with a large population of black holes that have mass segregated to the center of the cluster, and imagine that the system is embedded within a gas-rich interstellar medium (ISM). We will refer to the centrally confined population of black holes as the ``black hole subsystem'' and characterize the surrounding ISM with a gas density, $\rho_\infty$, and a sound speed, $c_\infty$. The star cluster can accumulate gas from the ISM, which was previously explored by \cite{naiman_2011}. In this previous work, the authors found that dense star clusters can harbor sizeable gas accumulations if the condition $\sigma/c_\infty \gtrsim 1$ is satisfied, where $\sigma$ is the velocity dispersion of the cluster \citep[see also][]{lin_2007,naiman_2009}. If we initially neglect the black hole subsystem and model the star cluster with a Plummer potential, the central density enhancement is then, 
\begin{equation}
    \rho_{\rm c} = \rho_\infty \left[1  + (\gamma - 1)\frac{\sigma^2}{c_\infty^2}\right]^{\frac{1}{\gamma - 1}}
\label{eq:density_enhancement},
\end{equation}
where $1 < \gamma < 5/3$ is the ratio of specific heats of the gas. The black hole subsystem will be embedded within this high density gas, from which it will accrete. To start, we will write down the ``Bondi-Hoyle'' accretion rate \citep{bondi_hoyle_1944}, which is the rate at which an isolated black hole accretes from the surrounding medium,
\begin{equation}
    \dot{M}_{\rm BH,\infty} = \frac{4{\pi}G^2M_{\rm BH}^2\rho_\infty}{(v_\infty^2 + c_\infty^2)^{3/2}},
\label{eq:mdot_bh}
\end{equation}
where $M_{\rm BH}$ is the black hole mass and $v_\infty$ is the velocity of the black hole with respect to the ambient gas. This rate is linearly proportionate to the local gas density, suggesting that a black hole at the center of the cluster should accrete at a rate enhanced by a factor $\rho_{\rm c}/\rho_\infty$. However, black holes can also accrete more efficiently when they exist in dense populations, such as the subsystem we are considering here. We studied this previously in \cite{kaaz_2019}, where we found that black holes accreting in dense populations should have their accretion rates boosted by a factor $N^\alpha$, where $N$ is the total number of black holes in the population and $0<\alpha<1$ depends on $\gamma$ and on the number density of the black holes. Together, this suggests that a black hole in the subsystem should accrete at a rate
\begin{equation}
    \dot{M}_{\rm BH,ss} = N^\alpha\left(\frac{\rho_{\rm c}}{\rho_\infty}\right)\dot{M}_{\rm BH,\infty}
    \label{eq:mdot_bhss}
\end{equation} 
As each black hole in the subsystem accretes, the infalling gas will produce an accretion luminosity, $L_{\rm BH}\approx \eta\dot{M}_{\rm BH,ss}c^2$, where $\eta$ describes the radiative efficiency of the accreted material. From Earth, this black hole population would be unresolved, and so we would instead observe the integrated accretion luminosity of the subsystem, $L_{\rm total} \approx NL_{\rm BH}$. Combined with our previous estimates, this results in the following relation,
\begin{equation}
L_{\rm total} \approx {\eta}N^{1+\alpha}\left(\frac{\rho_{\rm c}}{\rho_\infty}\right)\dot{M}_{\rm BH,\infty}c^2
\label{eq:lum_cusp}
\end{equation}
This luminosity can be significant and from Earth would likely be observed as an unresolved X-ray point source. To estimate its value, its important to first consider the astrophysical context where this type of accretion process can occur. In particular, we require the following criteria; (i) the host cluster is massive, (ii) the ambient gas is dense and sufficiently cool ($\sigma/c_\infty \gtrsim 1$), (iii) the host cluster has a mass segregated black hole population, and (iv) the cluster is not so young that feedback prevents gas from accumulating. YMCs satisfy the first criterion by definition, and typically reside in gas-rich galaxies with high star formation rates, so they typically satisfy the second criterion as well. The third and fourth criteria depend sensitively on the joint evolution of the black hole population and the cluster gas reservoir, which we have sketched in Figure \ref{fig:cartoon}. When the cluster is newly born ($\lesssim1\,{\rm Myr}$), stellar evolution has only just begun, and the cluster retains its natal gas. Over the course of the next $\sim5-50\,{\rm Myr}$, the massive stars in the cluster will evolve and die. During this period, stellar wind and supernova feedback efficiently drive out the natal gas from the cluster. Meanwhile, the black hole population will mass segregate, and form a distinct, centrally confined subsystem on roughly the two-body relaxation timescale ($\approx 50-100\,{\rm Myr}$). Once both the black holes have mass segregated and the majority of the massive stars have evolved and died, then the black holes can begin accreting within a sound crossing time ($\lesssim10\,{\rm Myr}$). The black holes will gradually be ejected from the cluster, and the black hole subsystem can exist for roughly $\approx 10$ relaxation timescales ($\approx 500\,{\rm Myr}-1\,{\rm Gyr}$). So, we expect that accreting black hole subsystems can only be harbored by YMCs with ages roughly in the range of $\approx 50\,{\rm Myr}-1\,{\rm Gyr}$. 

We now turn our attention to the nearby Antennae Galaxies, which host several clusters that are promising candidates for hosting accreting black hole subsystems. They have a large population of YMCs that has been observed extensively in the optical and infrared \citep{Whitmore1995,Whitmore_2010}, and while most are newly formed ($\lesssim10\,{\rm Myr}$), a significant fraction are in the intermediate age range that we are interested in ($\approx 50-500\,{\rm Myr}$). Also, the ISM of the Antennae Galaxies is rich in cool ($c_\infty\sim10-20\,{\rm km\,s^{-1}})$, relatively dense ($\rho_\infty\sim10^{-23}-10^{-22}\,{\rm g\,cm^{-3}}$) gas \citep{zhu_2003,gilbert_2007,Weilbacher2018}. For a cluster with velocity dispersion $\sigma \approx 40\,{\rm km\,s^{-1}}$ and an accreting subsystem of one hundred $\approx 20\,M_\odot$ black holes, the collective accretion luminosity of the black holes is roughly $L_{\rm total} \approx 10^{39}-10^{43}\,{\rm ergs/s}$ (Equation \ref{eq:lum_cusp}), where the precise value depends most strongly on the efficiency of cooling. This is extremely luminous, and if accreting black hole subsystems currently exist within the Antennae Galaxies, they should be readily observable. If our proposed scenario is viable, then the detection or non-detection of the luminous signal would provide strong constraints on the population of black holes in young stellar environments.

The characteristics of this potential signal depend sensitively on the hydrodynamics of the cluster environment, which we probe by performing a set of hydrodynamic simulations that explore the relevant parameter space. In Section \ref{sec:methods}, we review the hydrodynamic setup that we use and motivate our choice of initial conditions and parameters. In Section \ref{sec:results}, we investigate the results of our simulations and study how the calculated accretion rates depend on these parameters. We make predictions for the luminous signals of these systems in Section \ref{sec:discussion}, and consider the observational constraints that our predictions place on black hole subsystem populations in the Antennae Galaxies.

\section{Methodology}
\label{sec:methods}
In this section, we present the framework to contextualize our results, review our hydrodynamic setup, and provide motivations for the parameter space we chose to explore. 

\subsection{Characteristic Scales}
\label{subsec:scales}
Before proceeding, it's first necessary to introduce a few key quantities that will be referenced throughout this work. We will use the subscript ``c'' to refer to the cluster and ``ss'' to refer to the black hole subsystem. The most important quantity is the \textit{accretion radius}, $R_{\rm a}$, which refers to the impact parameter within which a mass can gravitationally focus the surrounding gas \citep{hoyle_lyttleton_1939,edgar_2004}. Because the mass scales we simulate span several orders of magnitude, so too do the length scales we are interested in. 
At large scales, the cluster potential is dominant, which causes gas to accumulate on length scales of the cluster accretion radius, 
\begin{equation}
    R_{\rm a,c} = \frac{2GM_{\rm c}}{v_{\infty}^2} = 20\,{\rm pc}\left(\frac{M_{\rm c}}{10^6\,M_\odot}\right)\left(\frac{v_\infty}{20\,{\rm km\,s^{-1}}}\right)^{-2},
\label{eq:ra_cluster}
\end{equation}
where $M_{\rm c}$ is the mass of the host cluster. At smaller scales, where the subsystem accretion processes occur, the potential of the black hole subsystem members begins to dominate. We can characterize these length scales using either the subsystem or the individual black hole accretion radii,
\begin{equation}
    R_{\rm a,ss} = \frac{2GNM_{\rm BH}}{v_{\infty}^2} = 0.04\,{\rm pc}\left(\frac{N}{100}\right)\left(\frac{M_{\rm BH}}{20\,M_\odot}\right)\left(\frac{v_\infty}{20\,{\rm km\,s^{-1}}}\right),
\label{eq:ra_bh}
\end{equation}
where $M_{\rm BH}$ is the mass of an individual black hole and $N$ is the total number of subsystem members. We also define the ratio of the subsystem mass to the host cluster mass to be
\begin{equation}
    m = \frac{NM_{\rm BH}}{M_{\rm c}}
\label{eq:mass_ratio}
\end{equation}
In general, the black hole subsystem potential is highly subdominant to the host cluster potential (i.e., $m \ll 1$). 

\subsection{Simulation Details}
\label{subsec:numerics}

For all simulations, we use version 4.5 of the grid-based adaptive-mesh-refinement (AMR) hydrodynamics code FLASH \citep{fryxell_2000}. We use a nearly identical setup as \cite{kaaz_2019}, which we refer the reader to for a detailed summary of our computational approach. In the passages that follow, we summarize the most salient features of our simulations:
\begin{itemize}
\item We assume that the self-gravity of the gas is negligible, and accordingly use a scale-free, dimensionless setup. The gas densities required for star formation are much higher than those required for the black hole subsystem to achieve significant accretion luminosities. In our cases of interest, the gas is unlikely to be cooler than $10\,{\rm km\,s^{-1}}$, and for gas concentrated within a core radius that is $\approx 0.5\,{\rm pc}$, the Jeans mass is always $\gtrsim10^4\,M_\odot$. The total gas mass within this region can only ever reach $10^4\,M_\odot$ in extreme circumstances.
\item We embedded our cluster within a Hoyle-Lyttleton "wind-tunnel" \citep{blondin_2012} domain, where a supersonic, uniform gaseous wind of density $\rho_\infty$ and velocity $\textbf{v}_\infty = v_\infty\hat{\textbf{x}}$ is incident on the cluster and is continuously replenished every timestep. In all simulations, we used a Mach number $\mathcal{M} = 2$ and a computational domain bounded by $[x_{\rm min},x_{\rm max}] = [-12,20]\,R_{\rm a,c}$, $[y_{\rm min},y_{\rm max}] = [z_{\rm min},z_{\rm max}] = [-22.4,22.4]\,R_{\rm a,c}$.
\item We represented our black hole subsystems as ensembles of $N$ uniformly distributed, equal-mass "accretors", with fixed radii $R_{\star}$ and discretized potentials with mass $M_{\rm BH}$. While this is a highly simplified distribution, simultaneously modelling the gravitational encounters between the accretors would be computationally expensive, and we don't expect the dynamical feedback or relative velocity of the accretors to significantly affect our results. Every timestep, we flattened the density and pressure within the boundary of each accretor, and used the removed mass to calculate their accretion rates. 
\item We represented our host cluster by using a Plummer potential with mass $M_{\rm c}$ and core radius $r_{\rm c}$ \citep{plummer_1911},
\begin{equation}
    \Phi_{\rm c} = \frac{GM_{\rm c}}{(r^2 + r_{\rm c}^2)^{1/2}}
\label{eq:plummer}
\end{equation}
\item The main difficulty of this work was simultaneously resolving the large-scale hydrodynamics on scales of $R_{\rm a,c}$ while also resolving accretion processes on scales of $\approx 10^{-2}\,R_{\rm a,BH}$. To alleviate this computational difficulty, we simulated solely the cluster potential at low resolutions until $20\,R_{\rm a,c}/v_{\infty}$, by when the large-scale gas accumulation reached steady state. We then activated the black hole subsystem and continued evolving our simulations for time $20\,R_{\rm a,ss}/v_\infty$. During this accretion phase, we dramatically increased the resolution such that the minimum cell size is $\delta_{\rm min} = 0.25\,R_{\rm \star}$.
\end{itemize}

\subsection{Cluster Environments}
\label{subsec:params}

We motivate our choices of initial cluster parameters, including properties of the BH subsystem, using results from the dynamical simulations of cluster evolution from
\cite{kremer_2019c}. In particular, we consider the results of their solar metallicity calculations for cluster ages between $50\,\,{\rm Myr}$ and $1\,\,{\rm Gyr}$. Our choices for various cluster properties are described below. 

\begin{itemize}
    \item The predicted black hole subsystem populations of the clusters we want to represent range from hundreds to thousands of black holes. However, due to computational limitations, we limit our simulated subsystem populations to $N = 4$, $16$ and $32$ black holes. In \cite{kaaz_2019} and Subsection \ref{subsec:scaling_results}, we provide scaling relations that allow us to extrapolate to higher $N$.
    \item Our simulations are performed in dimensionless Hoyle-Lyttleton (HL) units ($R_{\rm a,c} = 1,\, v_{\infty}=1$) such that our results are generalizable to a wide range of cluster environments. 
    \item In all simulations, we use a core radius of $r_{\rm c} = 0.125\,R_{\rm a,c}$, which corresponds to $r_{\rm c}=1\,{\rm pc}$ when $M_{\rm c} = 10^6\,M_\odot$. This also fixes the ratio of the velocity dispersion to the ambient sound speed to $\sigma/c_\infty = 4$, which is a critical parameter in determining the efficiency of collective accretion onto core potentials \citep{naiman_2011}.
    \item We also use a fixed subsystem-to-cluster mass ratio of $m = 1/2^6$. While these masses are greater than those found in realistic systems, we stress that due to the dimensionless nature of our calculations, the particular choice of black hole mass is unimportant as long as $m \ll 1$. 
    \item We fix the mean separation between subsystem members, which plays a key role in determining the efficiency of collective accretion by the subsystem \citep{kaaz_2019}, to $R_{\perp}/R_{\rm a,BH} = 0.5$. When $M_{\rm c} = 10^6\,M_\odot$, $v_{\infty} = 20\,{\rm km\,s^{-1}}$ and $N = 32$, this corresponds to a subsystem number density of $n \approx 10^7\,{\rm pc}^{-3}$, which is consistent with simulations of cluster evolution at solar metallicity \citep{kremer_2019c}.
\end{itemize}

\subsection{Equation of State}
\label{subsec:cooling}

The efficiency of cooling can either inhibit or facilitate high accretion rates. In this subsection, we briefly consider the role of radiative losses in cluster environments to contextualize our choices for the equation of state. We consider conditions appropriate to the ISM of the Antennae Galaxies, where the temperature is likely $\approx 10^4-10^5\,{\rm K}$ and the gas density is of order $\approx 10^{-22}-10^{-23}\,{\rm g\,cm^{-3}}$ \citep{zhu_2003,gilbert_2007,Weilbacher2018}, making the gas an optically thin plasma that is sensitive to line cooling and Bremhsstrahlung losses.  We characterize these cooling processes with an emissivity function, $\Lambda(T)$, which can be related to a radiative energy loss rate as,
\begin{equation}
    \dot{E}_{\rm rad} \approx \left(\frac{\rho_{\rm c}}{m_{\rm H}}\right)^2V\Lambda(T),
\label{eq:edot_rad}
\end{equation}
where $V$ is the estimated volume of the system (which scales approximately as $\approx  R_{\rm a,c}^3$) and $\rho_{\rm c}$ is the density of the accumulated material (Equation \ref{eq:density_enhancement}). To characterize the efficiency of this cooling, we must compare it to the relevant energy deposition rates in the cluster environment. Here, we consider the Hoyle-Lyttleton energy deposition rate,
\begin{equation}
    \dot{E}_{\rm HL} = \frac{2\pi G^2M^2_{\rm c}\rho_\infty}{v_{\infty}},
\label{eq:edot_hla}
\end{equation}
which results from the thermalization of the supersonic wind at the bow shock of the gas accumulation. In reality, the luminosity field of the host cluster and stellar wind feedback can can also deposit energy into the gas accumulation. For simplicity, we do not explicitly consider these heating sources here, but discuss their effects in Section \ref{sec:discussion}.

If $\dot{E}_{\rm rad} > \dot{E}_{\rm HL}$, then the system cools efficiently. Because $\dot{E}_{\rm rad} \propto R_{\rm a,c}^3$ and $\dot{E}_{\rm HL} \propto R_{\rm a,c}^2$, it is expected that as the cluster mass increases, cooling becomes more efficient. For YMCs in the Antennae Galaxies, it is generally true that $\dot{E}_{\rm rad} \gg \dot{E}_{\rm HL}$, indicating that the gas accumulated by these clusters is nearly isothermal. While for comparison we simulate accretion flows with both $\gamma = 5/3$ and $1.1$, this suggests that our $\gamma = 1.1$ simulations are better representations of realistic young cluster environments. For reference, we provide more quantitative cooling constraints in Figure \ref{fig:luminosityConditions}, which verify this assumption. 

\section{Results}
\label{sec:results}
In this section, we present the results of all production simulations. We will begin by examining the flow morphology and accretion rates of our adiabatic ($\gamma = 5/3$) simulations and consider the implications for cluster environments that cool inefficiently. Then, we will turn our attention to cluster environments that can cool efficiently, and present the calculations of our quasi-isothermal ($\gamma = 1.1$) simulations. Finally, we will study these results in the context of our previous work, and determine if the scaling relations provided in \cite{kaaz_2019} hold in more realistic cluster environments. 

\subsection{$\gamma = 5/3$}
\label{subsec:adiabatic_results}
We first consider the results of our adiabatic, $\gamma = 5/3$ simulation with $N=32$ black hole subsystem members. In general, we expect that a $\gamma = 5/3$ equation of state should strongly inhibit accretion, due to the dominant role pressure plays in an inefficiently cooling plasma. 

\begin{figure*}[bht]
    \centering
    \includegraphics[width=\textwidth]{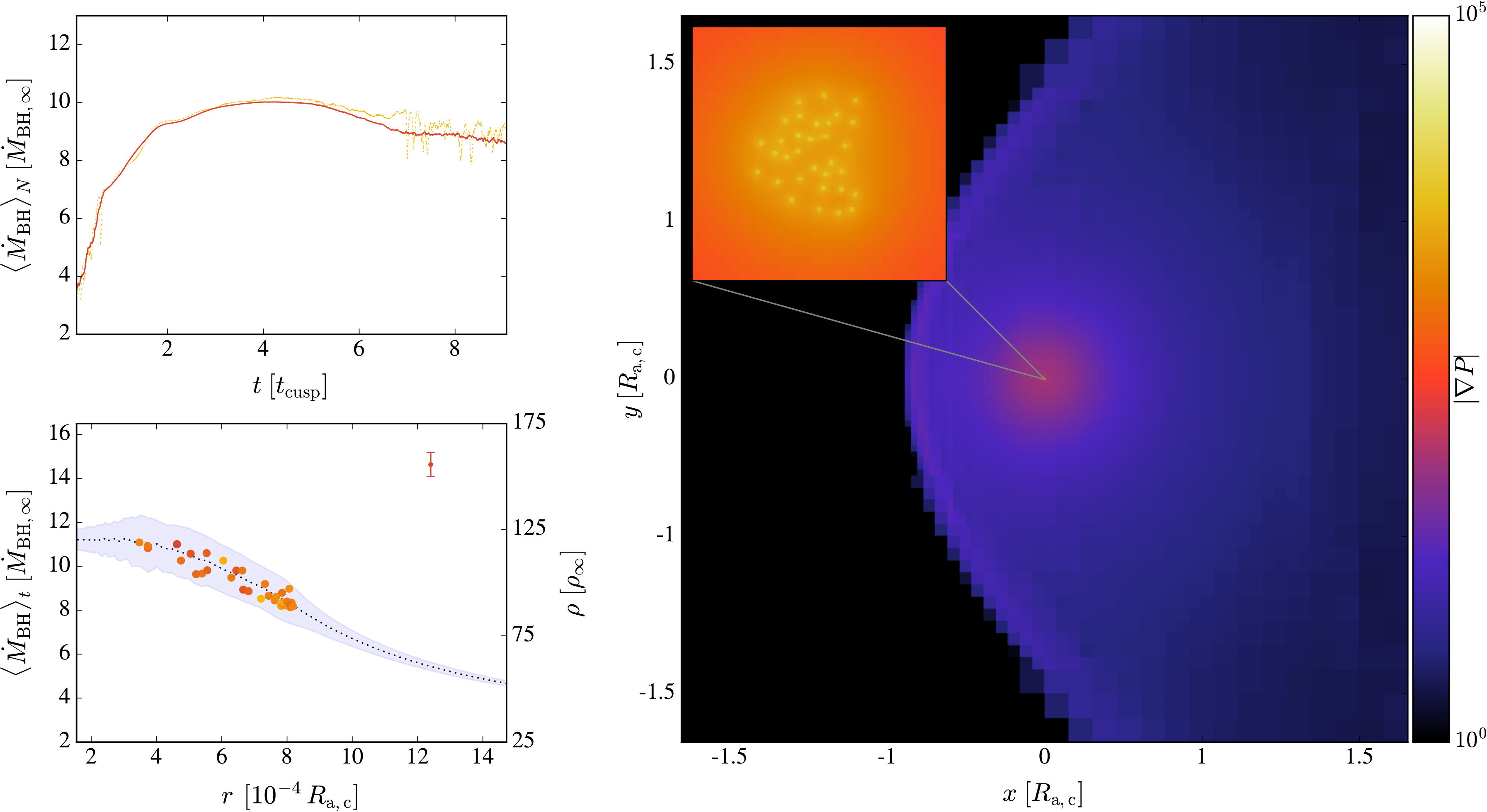}
    \caption{The accretion rates and flow characteristics are presented for our adiabatic $N=32$ simulation. \textbf{Right panel.} The projection of the pressure gradient of the flow is depicted within a $3\times3$ $R_{\rm a,c}$ box. The inset panel shows the innermost region of the simulation domain, exhibiting the local gas structure surrounding the individual subsystem members. \textbf{Upper left panel.} The number-averaged individual accretion rate of this simulation is plotted (red), normalized to its value in the ambient medium. The accretion rate for a randomly selected subsystem member is also depicted for reference (yellow). After time $\approx 4$ $t_{\rm cusp} = 4NR_{\rm a,BH}/v_\infty$, the accretion rate has reached steady state, achieving a modest $\approx 10\times$ enhancement from its ambient value. \textbf{Lower left panel.} The time-averaged individual accretion rates (left axis) are overplotted with the averaged density profile (right axis) as a function of radial position. The density profile is represented by the dotted black line, with the $\pm1$ standard deviation density contours enclosed by the shaded blue region. The accretion rate of each subsystem member is represented by a dot, where redder (yellower) colors indicate higher (lower) variability in the accretion rate. The average standard deviation for the accretion rates is plotted for reference in the upper-right corner.}
    \label{fig:adiabaticPanels}
\end{figure*}

In Figure \ref{fig:adiabaticPanels}, we depict the most important features of our $N=32$ adiabatic simulation. In the right panel, the magnitude of the pressure gradient is projected within a $3\times3\,R_{\rm a,c}$ cross-section of the simulation domain. The large-scale flow morphology is similar to what has been seen in previous cluster \citep{kaaz_2019,naiman_2011} and point mass \citep{blondin_2012} simulations; a stable, axisymmetric bow shock forms around the center-of-mass along the vector of the gaseous wind. The central region is embedded within a pocket of high-density material, which is insulated from the ram pressure of the supersonic wind due to the large stand-off distance of the bow shock. In the inset panel, we have zoomed in on a $6.4\times6.4\,R_{\rm a,BH}$ cross-section of the accreting subsystem. The subsystem is embedded within a hydrostatic envelope, with each individual subsystem member focusing the gas to higher densities in their local vicinity. 

In the upper-left panel of Figure \ref{fig:adiabaticPanels}, we plot the number-averaged accretion rates of the accreting subsystem as a function of time. The accretion rates are normalized to the numerical value of a single black hole accreting directly from the ambient medium, which we have calculated to be $\dot{M}_{\rm BH,\infty} = 0.49\,\dot{M}_{\rm HL}$ for $\gamma = 5/3$. Time is plotted in units of the subsystem crossing time, $t_{\rm cusp}=NR_{\rm a,BH}/v_{\infty}$, which is the time the ambient gaseous wind takes to cross one subsystem accretion radius. Surprisingly, the accretion rates of the subsystem members within the central density enhancement are only enhanced by a factor of $\lesssim10$ times their ambient values. This is seemingly in conflict with the analysis provided in Section \ref{sec:intro}, which suggested that the subsystem member accretion rates should be enhanced by a factor of $\approx{N}^{\alpha}\left(\frac{\rho_{\rm c}}{\rho_{\infty}}\right)\approx 50$. While the gas supply, and thus $\rho_{\rm c}$, is enhanced greatly from the ambient values, the gas is also virialized by the cluster potential, increasing the sound speed and temperature of the accumulated material. From Equation \ref{eq:mdot_bh}, $\dot{M}_{\rm BH}\propto c_{\rm s}^{-3}$, and so the high temperatures of the accumulated material tempers the resulting accretion rates. This emphasizes the role of cooling in facilitating high accretion rates; if the gas can cool efficiently, then $c_{\rm s}(r=0)\approx c_{\infty}$, allowing the gas to compress to drastically higher densities.

In the lower-left panel of Figure \ref{fig:adiabaticPanels}, we present the positional information of the accretion flow within the central, high-density region. We over-plot both the individual, time-averaged accretion rates (scatter points) and the averaged radial density profile (black dotted line and blue shaded region) as a function of radial position. In addition, we represent the degree of variability in the accretion rate of each subsystem member by the color of the scatter points, where lighter yellow and orange symbols indicate lower variability and darker red symbols indicate higher variability. In general, the accretion rates follow the density structure, and the member-by-member variance in the time-averaged accretion rates fall within the $\pm1$ standard deviation contours of the density profile (blue shaded region). Both the accretion and density profiles are relatively flat, with no appreciable pattern in the variability of the accretion rates. This is due to the stability of the hydrostatic envelope; there is minimal local variance within the subsystem, and each accretor is effectively interchangeable with one another. While this is expected within $\gamma = 5/3$ gas, in efficiently cooling flows pressure support is stripped away from the system, making the internal structure of the density enhancement unstable and subject to fluctuations. We explore the consequences of this in the next section, where we analyze the results of our quasi-isothermal, $\gamma = 1.1$ simulations. 

\begin{figure*}[bht]
    \centering
    \includegraphics[width=\textwidth]{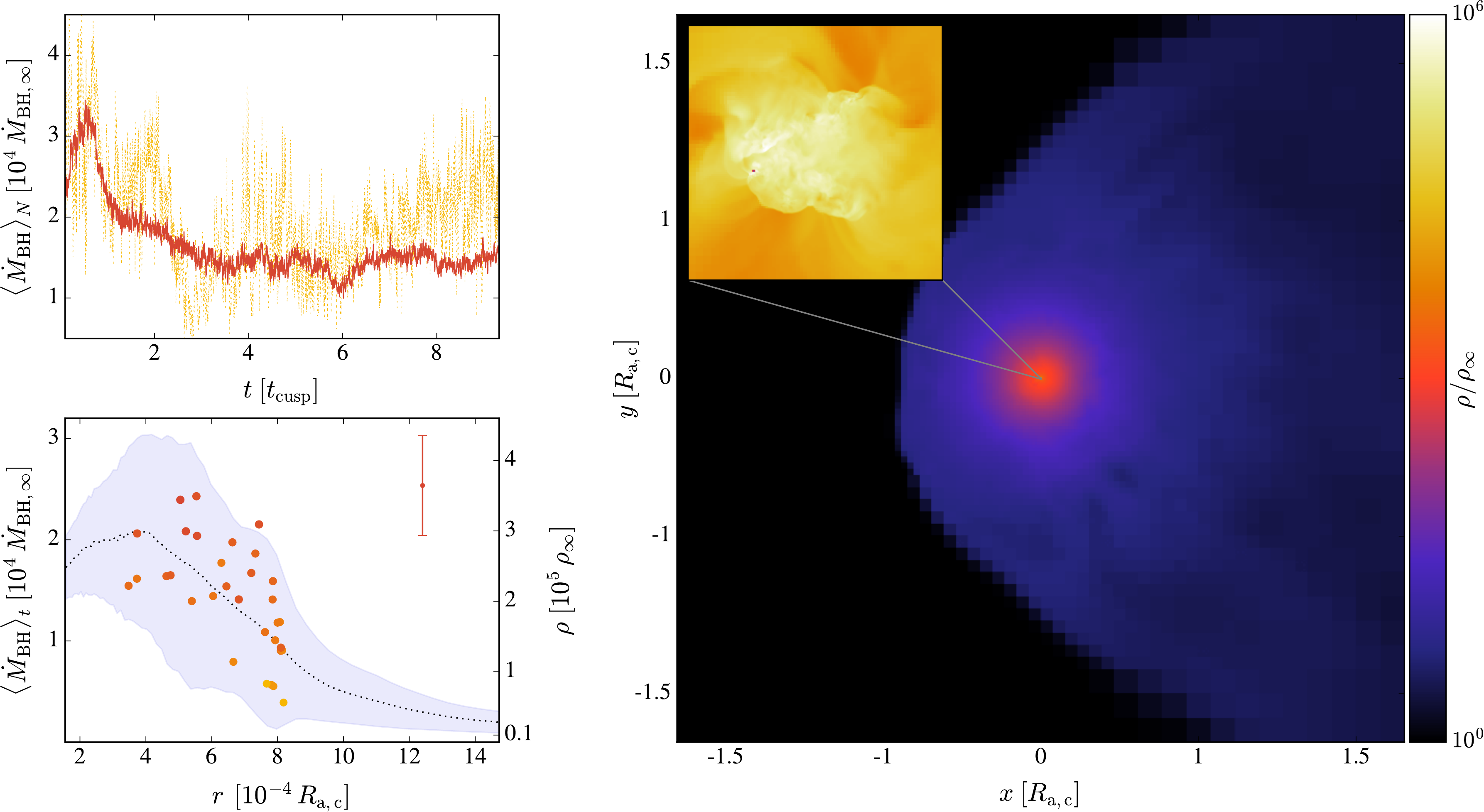}
    \caption{The accretion rates and flow characteristics are presented for our semi-isothermal ($\gamma = 1.1$) $N=32$ simulation. \textbf{Right panel.} The density of the flow is depicted within a $3\times3$ $R_{\rm a,c}$ slice. The inset panel shows the innermost regions of the simulation domain, depicting the large fluctuations that arise due to the inability of the sound speed in efficiently cooling gas to virialize. \textbf{Upper left.} The number-averaged individual accretion rate is plotted (red), normalized to its value in the ambient medium. The accretion rate for a randomly selected subsystem member is also depicted for reference (yellow). After time $\approx 4$ $t_{\rm cusp} = 4NR_{\rm a,BH}/v_\infty$, the accretion rate has reached steady state. \textbf{Lower left.} The time-averaged individual accretion rates (left axis) are overplotted with the azimuthally-averaged density profile (right axis) as a function of radial position. The density profile is represented by the dotted black line, with the $\pm1$ standard deviation contours enclosed by the shaded blue region. The accretion rate of each subsystem member is represented by a dot, where redder (yellower) colors indicate higher (lower) variability in the accretion rate. The average standard deviation for the accretion rates is plotted for reference in the upper-right corner.}
    \label{fig:isothermalPanels}
\end{figure*}

\subsection{$\gamma = 1.1$}
\label{subsec:isothermal_results}
We now explore the consequences of using a quasi-isothermal, $\gamma = 1.1$ equation of state on the accretion rates and flow morphology of our simulations. This represents the regime of efficient cooling, dramatically increasing the compressibility of the gas. We expect that the gas should be unable to virialize and that the stripped pressure support caused by radiative losses will allow much steeper density profiles to form, resulting in much higher accretion rates. 

We depict our $\gamma = 1.1$ results for our $N=32$ simulation in Figure \ref{fig:isothermalPanels}. In the right panel, we illustrate the large-scale flow morphology by depicting the density of a $3\times3\,R_{\rm a,c}$ slice of our computational domain, centered on the cluster center-of-mass. Similarly to our adiabatic simulations, the main features of Hoyle-Lyttleton accretion are present, including a central density enhancement and a bow shock. While the bow shock exhibits small-scale asymmetries, it is largely stable. This is not generally the case for isothermal simulations of Hoyle-Lyttleton flow; often, the heightened compressibility of the accumulated material causes the stand-off distance of the bow shock to shrink, and if it becomes smaller than the core radius a stable bow shock cannot form \citep{naiman_2011}. However, this threshold is sensitive to the choice of core radius, Mach number of the gaseous wind, and the amount of material accreted. If we use Equation 19 of \cite{naiman_2011} to determine this threshold analytically, we find that $r_{\rm s} \approx 1.5\,R_{\rm a,c}$, which is greater than our choice of $r_{\rm c} = 0.125\,R_{\rm a,c}$ and consistent with our results.

In the inset panel of Figure \ref{fig:isothermalPanels}, we depict the gas density of a $6.4\times6.4\,R_{\rm a,BH}$ slice of the flow, within which the accretion subsystem is embedded. Due to radiative losses, the gas is unable to virialize, resulting in a high-density, time-dependent envelope of converging material that continuously feeds the accreting cusp. The drastically enhanced densities in this envelope allow the cusp members to accrete at very high rates, as depicted in the upper-left panel of Figure \ref{fig:isothermalPanels}. Here, we again plot the number-averaged accretion rates normalized to their ambient value (which for $\gamma = 1.1$ we have calculated to be $\dot{M}_{\rm BH,\infty}=1.04\dot{M}_{\rm HL}$) as a function of time, in units of the cusp crossing time $t_{\rm cusp}$. The presented accretion rates are remarkably higher than their ambient values, achieving steady states values of $>10^{4}\,\dot{M}_{\rm BH,\infty}$. Additionally, the early, transient behavior of $\langle\dot{M}_{\rm BH}\rangle_N$, is different from both our corresponding adiabatic simulation (Figure \ref{fig:adiabaticPanels}) and previous works \citep{kaaz_2019,blondin_2012}. Instead of gradually reaching steady-state, the accretion rate sharply increases, and then decreases to its steady-state value. We generally frame our analysis of the accreting cusp in the context of BHL accretion, except that the ambient density $\rho_\infty$ is replaced by the central, enhanced density $\rho_{\rm c}$. While this is a good approximation for our adiabatic simulations, in which the subsystem is embedded within a hydrostatic envelope, it becomes less appropriate for our isothermal simulations. This is for two reasons; first, even in the absence of the accreting subsystem the gas could never virialize, meaning that the density enhancement could never achieve steady-state, and the flow structure is not similar to the ambient conditions at infinity. Second, in the isothermal simulation, the total accretion rate of the subsystem becomes a significant fraction ($\approx10\%$) of the total gas supply to the cluster (estimated to be $\approx\dot{M}_{\rm HL,c}$, the Hoyle-Lyttleton accretion rate of the cluster). This causes the accreting subsystem to impact the structure of the accumulated material, making the comparison to canonical BHL less tenable, and resulting in the early transient spike in the accretion rate. 
In the lower-left panel of Figure \ref{fig:isothermalPanels}, we consider the time-averaged accretion rates of our subsystem members (scatter points), the variability in their accretion rates (color of the scatter points), and the density structure of the subsystem (black dotted line and blue shaded region). This can be directly compared to the analogous panel of Figure \ref{fig:adiabaticPanels}, where it is immediately apparent that there is much larger positional dependence within the subsystem of the $\gamma = 1.1$ simulation, and it remains true that the member-by-member variation in the time-averaged accretion rates follows the variation in the density structure. What is more striking is that there is now a much stronger positional dependence on the variability in the accretion rate; the outermost subsystem members have both much lower accretion rates and much lower variability, while the innermost subsystem members have higher accretion rates and higher variability. 

In the subsection that follows, we will attempt to bridge our results to canonical studies of BHL accretion. In particular, we will consider the viability of studying the behavior of accretors within a cluster density enhancement using the framework of BHL accretion, trading the usually prescribed ambient density $\rho_\infty$ with the enhanced density.

\subsection{Scaling relations}
\label{subsec:scaling_results}
In \cite{kaaz_2019}, we focused on developing physical intuition for collective BHL accretion in dense stellar environments. Our previous investigation was limited to small populations of $N \leq 32$ accretors, making the extrapolation to astrophysical stellar populations uncertain. The incorporation of the core potential in the simulations presented here allows us to test our intuition in a more realistic environment, where the potential of an individual cluster member is highly subdominant to the aggregate cluster potential. 

For each equation of state ($\gamma = 1.1$ and $5/3$), we seek to determine the scaling relation between the accretion rates of the subsystem members and the total subsystem population, 
\begin{equation}
    \dot{M}_{\rm BH} = \dot{M}_{N=1}N^{\alpha},
\label{eq:mdot_scaling}
\end{equation}
where $\dot{M}_{N=1}$ is our calculated accretion rate for a subsystem composed of a single black hole. In canonical BHL accretion, we expect the parameter $\alpha$ to be bounded between $0$ and $1$ (see Section 2 of \cite{kaaz_2019}). For $\gamma = 5/3$ and $1.1$, we calculate $\dot{M}_{N=1} = 2.21\,\dot{M}_{\rm BH,\infty}$ and $419\,\dot{M}_{\rm BH,\infty}$, respectively. Here, it is important to emphasis that $\dot{M}_{\rm BH,\infty}$ and $\dot{M}_{N=1}$ are distinct quantities; $\dot{M}_{\rm BH,\infty}$ refers to the accretion rate of a single black hole in the ambient medium, while $\dot{M}_{N=1}$ refers to the accretion rate of a single black hole embedded within the core potential. It is remarkable that for $\gamma = 5/3$, the accretion rate of a single black hole is only marginally enhanced from its ambient value, despite being embedded within the central high-density envelope. The reason for this is that the virialization of the gas by the cluster potential enhances the sound speed in the density enhancement, which keeps the accretion rate of the black holes low despite the large gas supply. When $\gamma = 1.1$, the gas can cool efficiently, preventing the sound speed from thermalizing. This and the removal of pressure support from the density enhancement in efficiently cooling systems is responsible for the large difference in calculated accretion rates between $\gamma = 1.1$ and $\gamma = 5/3$ simulations. 

\begin{figure}
    \centering
    \includegraphics[width=\textwidth]{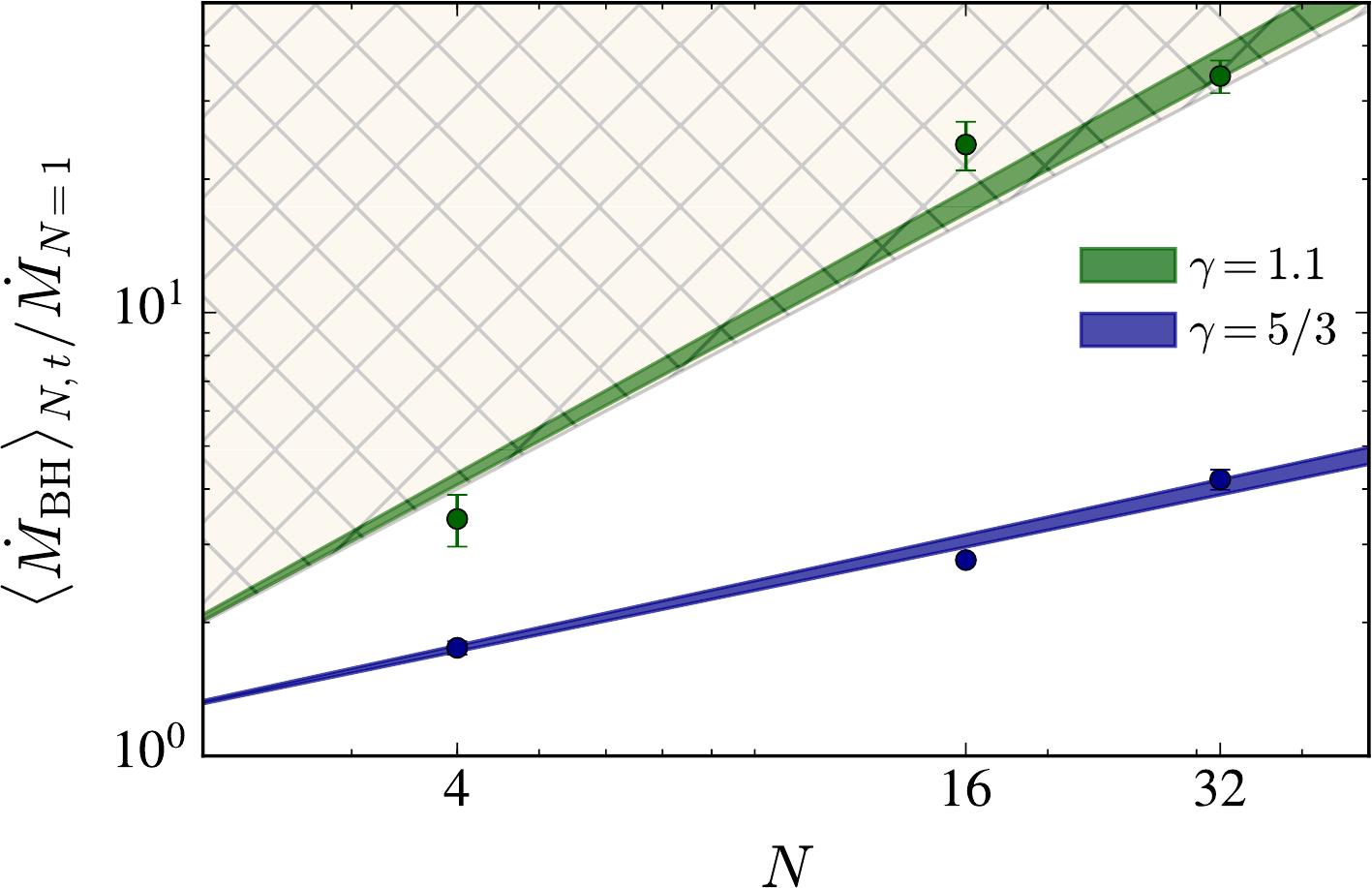}
    \caption{We depict the time- and number-averaged accretion rates of black hole subsystem members in both our $\gamma = 1.1$ and $\gamma = 5/3$ simulations. We have normalized to the accretion rate of a single subsystem member embedded within the host cluster gas accumulation, $\dot{M}_{N=1}/\dot{M}_{\rm BH,\infty} = 2.21$ and  $419$ for $\gamma = 5/3$ and $1.1$, respectively. We fit for the $\alpha$ parameter in Equation \ref{eq:mdot_scaling}, finding $\alpha = 0.40\pm0.01$ for $\gamma = 5/3$ and $1.04\pm0.02$ for $\gamma = 1.1$. The hatched region depicts the constraint $\alpha \lesssim 1$.}
    \label{fig:scalingRelation}
\end{figure}

In Figure \ref{fig:scalingRelation}, we plot the number- and time-averaged accretion rates of each simulation, normalized to their respective values of $\dot{M}_{N=1}$, as a function of $N$. We also provide our fit for the $\alpha$ parameter in Equation \ref{eq:mdot_scaling}, which we calculate to be $\alpha = 0.40\pm0.01$ and $1.04\pm0.02$ for $\gamma = 5/3$ and $\gamma = 1.1$, respectively. These results can be compared directly to Figure 11 of \cite{kaaz_2019}, where the same fits were performed for an identically-distributed system in the ambient medium rather than in the core potential density enhancement. Both analyses result in very similar values of $\alpha$ for a given $\gamma$. 

While this similarity suggests that our accreting subsystems behave analogously to a cluster accreting directly from the ambient medium, there are some limitations to this comparison. In canonical BHL accretion we assume that accretion occurs in an infinite, uniformly dense medium, but while the central density enhancement in our simulations is symmetric, it is not uniform. This can affect our accretion rates depending on how efficiently the subsystem can deplete material from the density enhancement. When $\gamma = 5/3$, the subsystem is embedded in a hydrostatic envelope and accretes at a low rate (Figure \ref{fig:adiabaticPanels}), only marginally perturbing the flow. The in-flow of material is steady, and so the gas supply is effectively uniform. However, when $\gamma = 1.1$, the envelope is both unstable and the subsystem accretes at a high rate, perturbing the dynamics of the flow. The perturbation depends on the total subsystem accretion rate, which depends on $N$. At higher $N$, although the individual accretion rates of subsystem members are enhanced, the subsystem as a whole accretes less efficiently. This is a consequence of our dimensionless setup; because we fix the subsystem mass, as we increase $N$ we split the subsystem mass into smaller components. As the subsystem depletes more of the accumulated material at lower $N$, the effective density decreases because of the finite density enhancement. It is to this effect that we attribute the slightly super-linear value of $\alpha$ for $\gamma = 1.1$; at lower $N$, the subsystem overall accretes more, lowering the effective density of the surrounding region. It is important to note that our choice of subsystem-to-cluster mass ratio, $m=1/2^6$, likely over-estimates that of real systems, and thus the effects of this perturbation are likely less robust in realistic cluster environments. Another difference in the comparison to canonical BHL is that Hoyle-Lyttleton flow is axisymmetric, while the central density enhancement is spherically symmetric, and the subsystem accretion is more similar to Bondi accretion \citep{bondi_1952}. The close match between the scaling relations presented in Figure \ref{fig:scalingRelation} and those in \cite{kaaz_2019} suggest that while the different geometries in Bondi and Hoyle-Lyttleton accretion will affect the amount of gas supplied, the physics of collective accretion will be unchanged. 

\section{Discussion}
\label{sec:discussion}

In the preceding sections, we have argued that black hole subsystems in young, dense cluster environments can produce large accretion luminosities provided that the quantity $\sigma/c_\infty$ is greater than unity. We have supported these arguments with hydrodynamic simulations, which have demonstrated that given that the accumulating gas can cool efficiently, black hole subsystems can accrete at very high rates. We will now build the connection from our results to their luminous signals, discussing observational prospects and such observations may place upon realistic cluster environments.

\subsection{The luminous signal of the accreting subsystem}

\begin{figure}
    \centering
    \includegraphics[width=\textwidth]{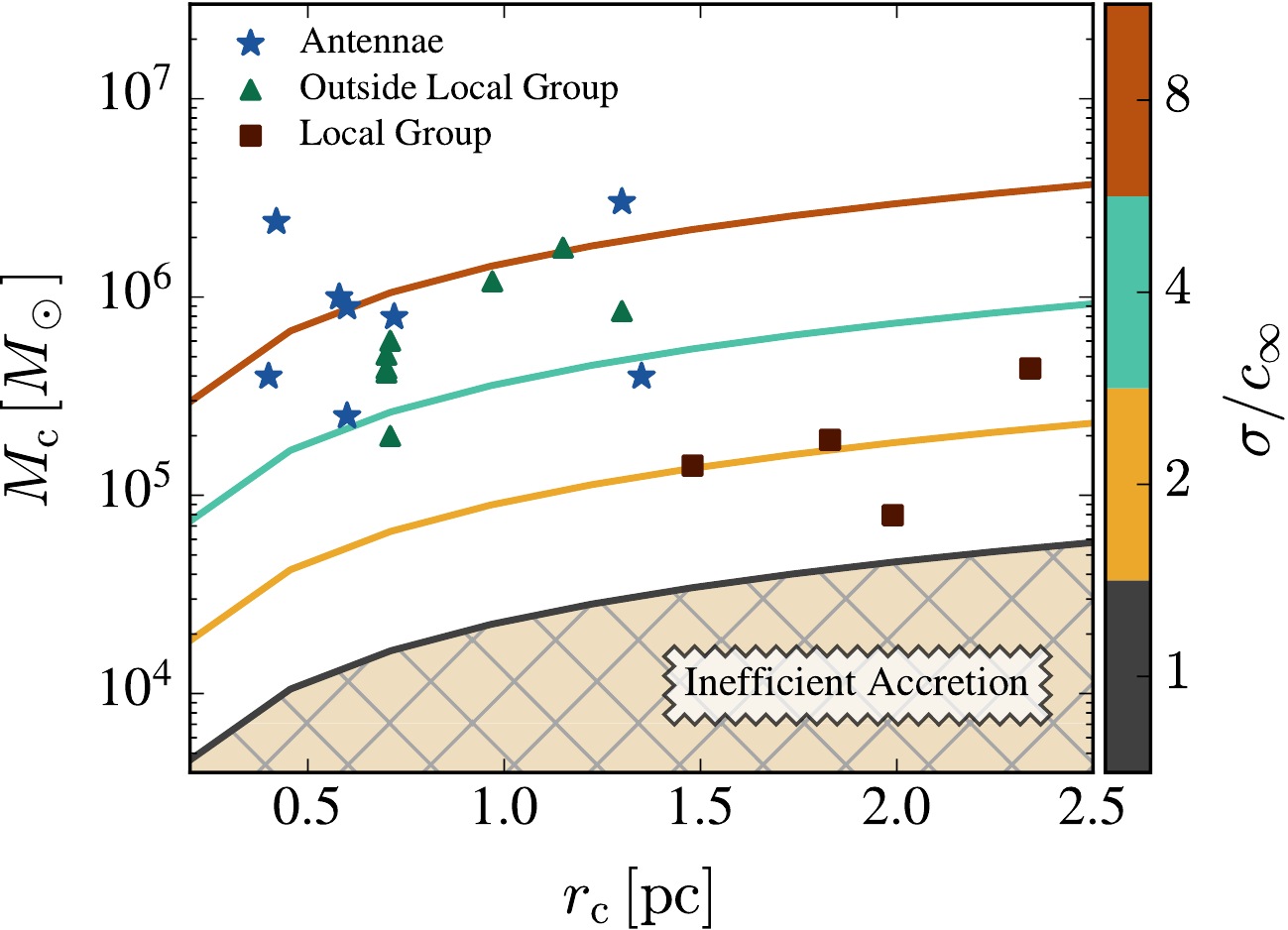}
    \caption{Cluster mass $M_{\rm c}$ is plotted as a function of core radius $r_{\rm c}$ for different values of $\sigma/c_\infty$, which determines the magnitude of the cluster gas accumulation. Each curve is calculated for sound speeds of $c_\infty = 10\,{\rm km\,s^{-1}}$, typical of the Antennae galaxies \citep{gilbert_2007,Weilbacher2018}. Clusters are only expected to accrete collectively if $\sigma/c_\infty \gtrsim 1$, which is reflected by the hatched region, where accretion is inefficient at $c_\infty = 10\,{\rm km\,s^{-1}}$. We have also plotted example YMCs from the Antennae complex as a reference \citep{zwart_2010}.}
    \label{fig:clusterEnvirons}
\end{figure}

Here, we infer accretion luminosities from our $\gamma = 1.1$ simulations presented in Section \ref{subsec:isothermal_results}. As a reminder, our simulations were performed in dimensionless HL units, such that the choice of ambient gas properties $\rho_\infty$ and $c_\infty$ are arbitrary. This formalism is advantageous as it allows us to rescale our results to different cluster environments.
Additionally, if the ratio of subsystem to cluster mass, $m\equiv NM_{\rm BH}/M_{\rm c}$, is much less than unity, then the subsystem negligibly perturbs the gas accumulation and the choice of black hole mass $M_{\rm BH}$ is also arbitrary. We also emphasize that the central density enhancement harbored by our cluster environment is set uniquely by the parameters $\sigma/c_\infty$ and $\gamma$ (Equation \ref{eq:density_enhancement}).
If we are to extrapolate our results to realistic cluster environments, it's crucially important that we choose the correct value of $\sigma/c_\infty$ to represent our system. From the definition of $\sigma^2\equiv GM_{\rm c}/r_{\rm c}$, we can write the following formula to consider how the cluster mass $M_{\rm c}$ behaves at fixed $\sigma/c_\infty$ as a function of core radius and ambient sound speed,
\begin{equation}
    M_{\rm c} = \left(\frac{\sigma}{c_\infty}\right)^2r_{\rm c}c_\infty^2/G
\label{eq:cluster_environs}
\end{equation}
In Figure \ref{fig:clusterEnvirons} we plot this relation for different values of $\sigma/c_\infty$. The solid curves in this figure assume that the ambient sound speed is $c_\infty = 10\,{\rm km\,s^{-1}}$, typical of the Antennae galaxies \citep{gilbert_2007,Weilbacher2018}. Collective accretion is only expected to occur if $\sigma/c_\infty\gtrsim1$ \citep{naiman_2011}; the hatched region in Figure \ref{fig:clusterEnvirons} indicates where collective accretion is not expected. A sample of YMCs with dynamical mass measurements from (i) the Antennae galaxies, (ii) beyond the Local Group (excluding the Antennae), and (iii) the Local Group \citep{zwart_2010} are also labeled for reference. These observed cluster parameters are generally above the $\sigma/c_\infty = 4$ curve, indicating that unless the sound speed is much higher than our assumed value of $10\,{\rm km\,s^{-1}}$, these clusters will accrete at least as efficiently as our results suggest.  Even under this more conservative assumption, the Antennae galaxies have several YMCs for which we would still be underestimating $\sigma/c_\infty$ and thus the associated gas accumulation. 

We will now infer subsystem luminosities from our results for various contexts. 
The subsystem luminosity is fully defined by the following parameters: the number of accretors $N$, the average black hole mass $M_{\rm BH}$, the exponent $\alpha$ (Equation \ref{eq:mdot_scaling}), the ambient gas density $\rho_\infty$, the ambient sound speed $c_\infty$, the velocity ratio $\sigma/c_\infty$ and the cluster Mach number $\mathcal{M}$. The appropriate value of $\mathcal{M}$ is unclear, but is likely of order unity, and we have assumed it to be $2$ in our analysis. The results of Section \ref{subsec:scaling_results} suggest that for an efficiently cooling gas accumulation, $\alpha = 1$ is appropriate, allowing us to use Equation \ref{eq:mdot_scaling} to extrapolate subsystem accretion rates to different values of $N$. We numerically calculated the quantity $\dot{M}_{N=1}/\dot{M}_{\rm BH,\infty} = 419$ for $\gamma = 1.1$ and $\sigma/c_\infty = 4$, which is how much the accretion rate of a single black hole gets enhanced relative to its ambient Bondi-Hoyle rate when it resides in its host cluster. This is equal to the dimensionless density enhancement $\rho_{\rm c}/\rho_\infty$ defined in Equation \ref{eq:density_enhancement} and is the only quantity dependent on $\sigma/c_\infty$. We can then write down the luminosity of the accreting subsystem as,
\begin{equation}
    L_{\rm total} = \eta N^{\alpha+1}\left[\frac{\dot{M}_{N=1}(\sigma/c_\infty,\gamma)}{\dot{M}_{\rm BH,\infty}}\right]\frac{4\pi G^2M^2_{\rm BH}\rho_\infty}{\mathcal{M}^3c_\infty^3}c^2
    \label{eq:lum_cusp_numerical}
\end{equation}

\begin{figure}
    \centering
    \includegraphics[width=\textwidth]{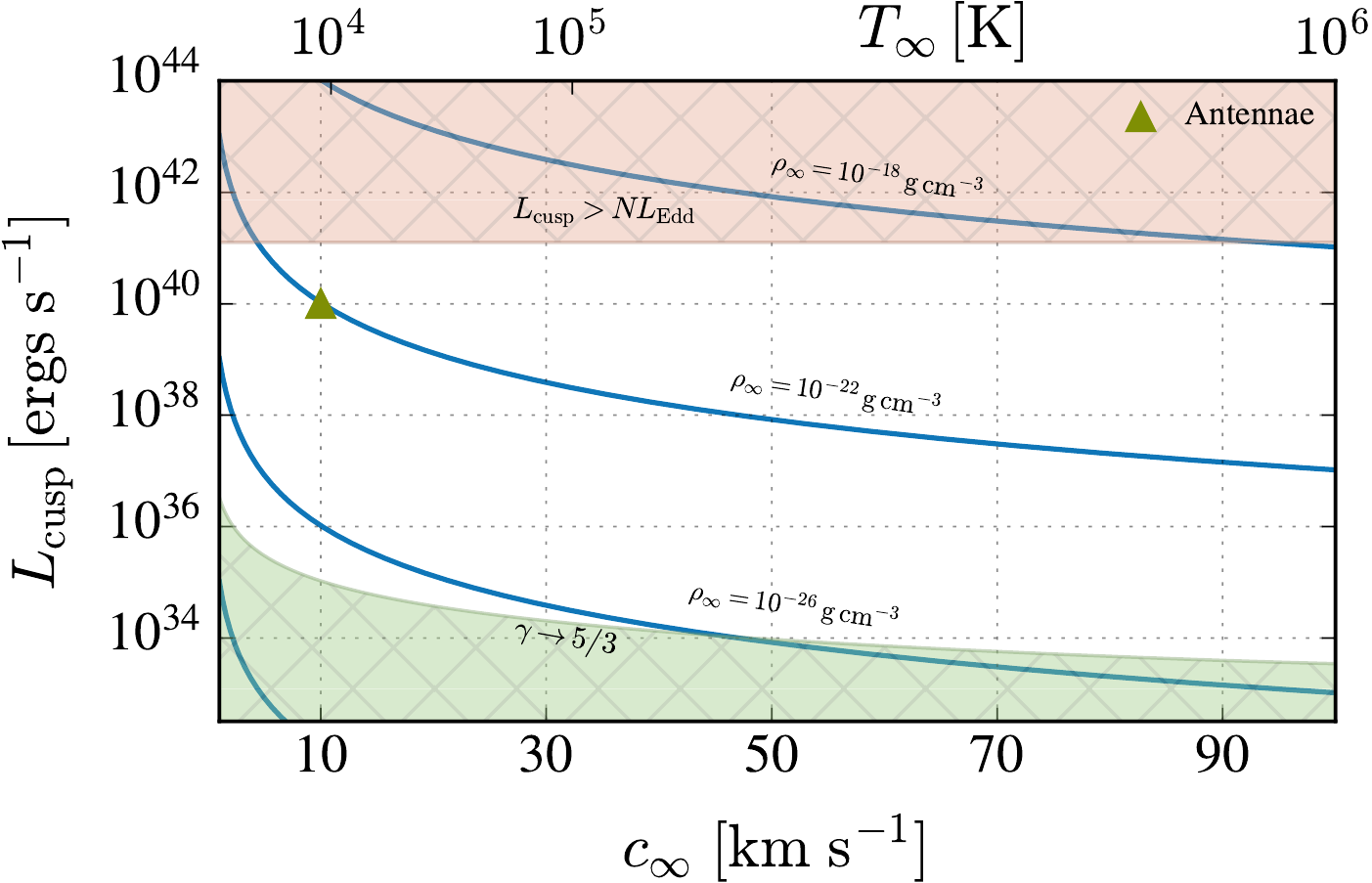}
    \caption{The subsystem luminosity is calculated for different values of $c_\infty$ and $\rho_\infty$, using Equation \ref{eq:lum_cusp_numerical}. We performed these calculations by using the results of our $\gamma = 1.1$ simulations, where the ratio of the velocity dispersion to the ambient sound speed is $\sigma/c_\infty = 4$. The precise values of the subsystem-cluster system are assumed to be: $\eta = 0.01$, $M_\odot$, $N=100$ and $M_{\rm BH} = 10\,M_\odot$. We also provided constraints for the Eddington limit of our accreting subsystem members and for the regime in which cooling ceases to be efficient, where to calculate the cooling constraint we assumed that $M_{\rm c} = 10^6\,M_\odot$. The triangular marker indicates the value of $L_{\rm cusp}$ appropriate to the ISM of the Antennae Galaxies, which we calculate to be $\approx10^{40}\,{\rm ergs\,s^{-1}}$.}
    \label{fig:luminosityConditions}
\end{figure}

This can be immediately compared to Equation \ref{eq:lum_cusp}, where the parameter representing the density enhancement in the host cluster is traded for the numerically determined quantity $\frac{\dot{M}_{N=1}}{\dot{M}_{\rm BH,\infty}}$. In Figure \ref{fig:luminosityConditions}, we plot this expression for our $\gamma = 1.1$,  $\sigma/c_\infty = 4$ simulations for different values of $\rho_\infty$ and $c_\infty$, where we have set $N=100$ and $M_{\rm BH} = 10\,M_\odot$. We emphasize that these choices of subsystem parameters are conservative, as solar metallicity cluster simulations typically predict subsystem populations of $\approx 10^3$ black holes of average mass $\approx 15\,M_\odot$ at early times \citep{kremer_2019c}. In addition to the luminosity contours for different ambient conditions, we have plotted constraints corresponding to the Eddington limit of the subsystem and to ambient properties where gas cools inefficiently ($\gamma \rightarrow 5/3$). The cooling constraint is provided by the condition $\dot{E}_{\rm HL}/\dot{E}_{\rm rad} = 1$ (Equations \ref{eq:edot_hla} and \ref{eq:edot_rad}). This relation depends on the total cluster mass, which we have assumed to be $M_{\rm c} = 10^6\,M_\odot$. When converting our accretion rates to luminosities, we have also conservatively estimated the radiative efficiency to be $\eta = 0.01$, due to the spherical geometry that the inflowing material exhibits on scales of $\sim R_{\rm a,ss}$.

At conditions similar to that of the Antennae complex (labeled by the triangular marker in Figure \ref{fig:luminosityConditions}), the implied accretion luminosity is significant, and the black holes accrete at an Eddington ratio of about $\approx 0.1$. As many of our parameter choices are conservative, the true accretion rate is potentially higher. This suggests that radiative feedback may be dynamically important in these systems. While hydrodynamic simulations with radiative transport are necessary to fully understand the importance of feedback, it is expected that this will cause black hole subsystems to accrete with some duty cycle which will reduce the number of accreting subsystems that can be observed at a given time. Regardless of the resulting effect on the statistics of luminous subsystems, the large number of YMCs observed in the Antennae Galaxies and similar systems indicate that if these systems have sizeable black hole subsystems and sufficiently dense embedded gas, we should be able to observe them. 

\begin{figure}
    \centering
    \includegraphics[width=\textwidth]{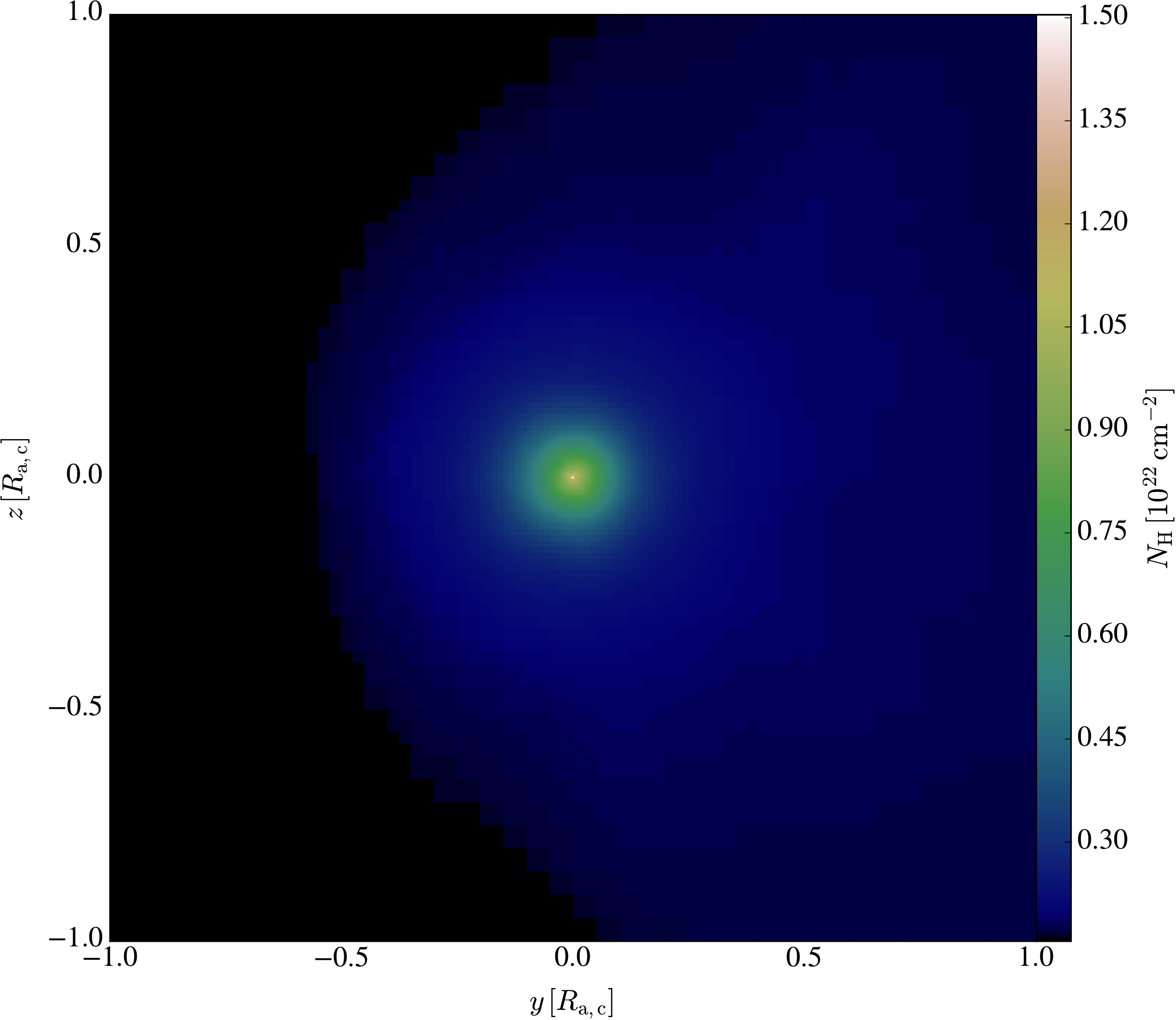}
    \caption{Here, we plot the Hydrogen column density in a $1\times1\,R_{\rm a,c}$ box for our $\gamma = 1.1$, $N=32$ simulation. We have assumed properties of the cluster environment such that  $N_{\rm H} \propto \frac{\rho_\infty}{10^{-24}\,{\rm g\,cm^{-3}}}\frac{M_{\rm c}}{10^6\,M_\odot}\left(\frac{c_\infty}{15\,{\rm km\,s^{-1}}}\right)^{-2}$.}
    \label{fig:column_density}
\end{figure}

Throughout this work, we have assumed that the cluster gas accumulation is optically thin and that the X-ray luminosity produced by the inner accretion flow readily escapes. This assumption deserves attention, because in cases where the density is significant, the total luminosity of the accreting subsystem can be impeded. In Figure \ref{fig:column_density}, we depict the Hydrogen column density for our $\gamma = 1.1$, $N=32$ simulation, where we have scaled our results to $\rho_\infty = 10^{-24}\,{\rm g\,cm^{-3}}$, $M_{\rm c} = 10^6\,M_\odot$, and $c_\infty = 15\,{\rm km\,s^{-1}}$. At these values, the Hydrogen column density reaches values of $\sim1.5\times10^{22}\,{\rm cm^{-2}}$. If we assume higher densities of $\rho_\infty=10^{-22}\,{\rm g\,cm^{-3}}$, and a cooler medium with sound speed of $c_\infty=10\,{\rm km\,s^{-1}}$, the maximum column density can reach $\approx 3\times10^{24}\,{\rm cm^{-2}}$. This corresponds to the characteristic gas conditions assumed for the Antennae galaxies (e.g., Fig. \ref{fig:luminosityConditions}). Once $N_{\rm H}$ is comparable to the inverse of the Thomson cross section $\sigma_{T}^{-1} \approx 1.5\times10^{24}\,{\rm cm}^{-2}$, the optical depth of Compton scattering becomes order unity, and the flow is by definition Compton-thick. If the column density remains below $\lesssim10^{25}\,{\rm cm}^{-2}$, then the flow is only ``mildly'' Compton-thick, but if the column density is any higher then the radiation will be significantly attenuated by Compton scattering. This suggests that in the Antennae galaxies, radiation is likely to be mildly attenuated by Compton scattering. Since the attenuation is mild, it wont significantly alter our results, and for the remainder of this work we will assume that the X-ray radiation escapes the accreting subsystem unimpeded. In the following section, we will combine our predicted luminosities with X-ray observations in the Antennae galaxies to provide constraints on YMC subsystem populations.

\subsection{Constraints imposed on the Antennae galaxies}
\label{subsec:antennae}

\begin{figure}
    \centering
    \includegraphics[width=\textwidth]{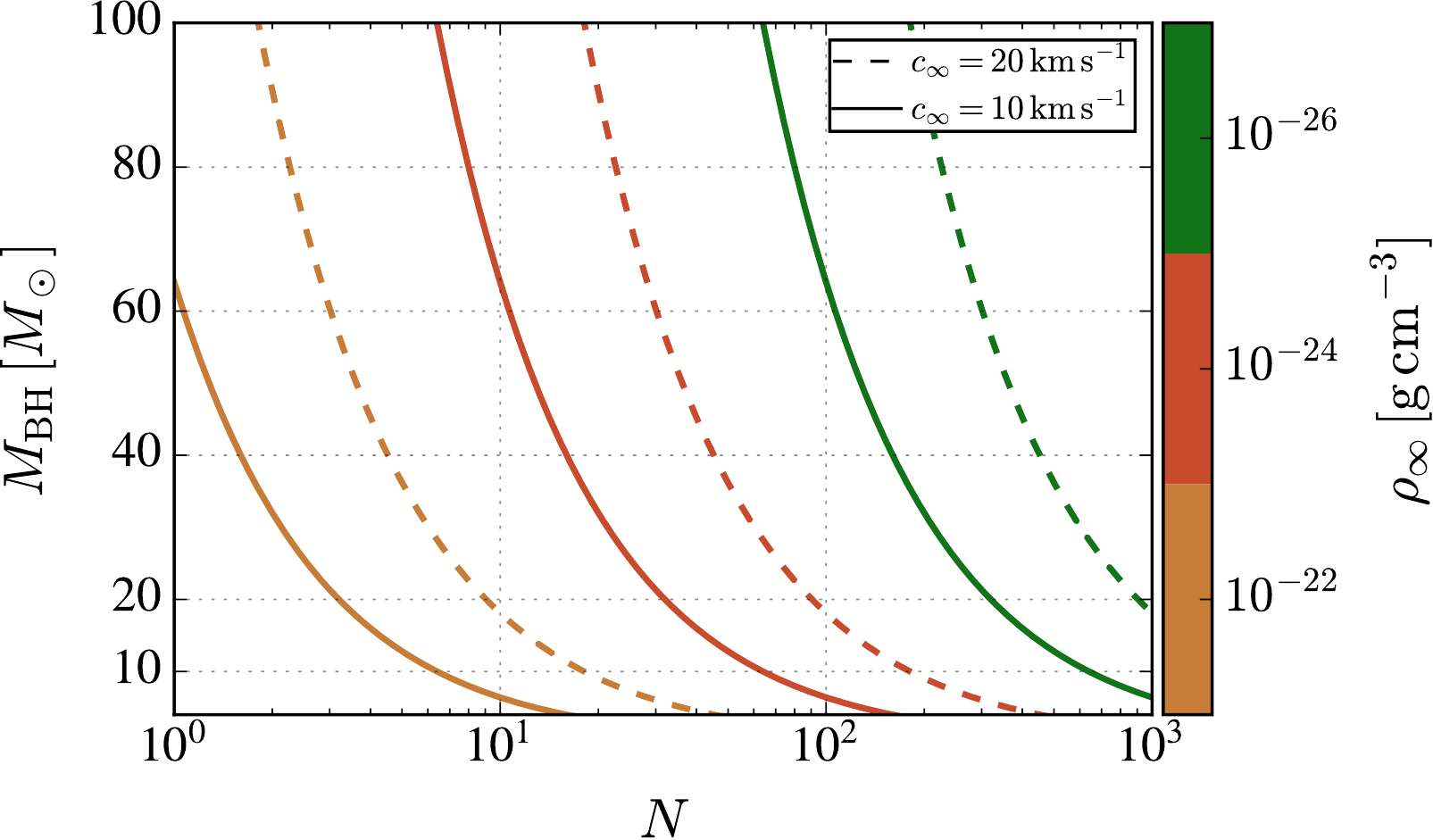}
    \caption{We depict constraints on the size of subsystems in YMCs where the condition $\sigma/c_\infty = 4$ is satisfied, assuming a limiting luminosity of $\approx 5\times10^{37}\,{\rm ergs\,s^{-1}}$ for the Antennae galaxies. We plot this relation for various ambient densities (distinguished by color), and for sound speeds of $10$ and $20\,{\rm km\,s^{-1}}$ depicted by solid and dashed lines, respectively. For a given density and sound speed, the $M_{\rm BH}-N$ parameter space is constrained to regions to the left of each curve.}
    \label{fig:cuspConstraint}
\end{figure}

We will now use our predicted accretion luminosities to place constraints on black hole subsystem populations in YMCs hosted by the Antennae Galaxies. The luminous signal of the accreting subsystem would most resemble that of so-called ultra-luminous X-ray sources (ULXs), which are abundant in the Antennae Galaxies \citep{zezas_2002}. The work of \cite{poutanen_2013} is most relevant to our interests because it studies the spatial correlations between the positions of YMCs and ULXs. They found that, in general, ULXs were associated but consistently offset from YMCs. This suggests that these ULXs are accreting binary systems that were ejected from their host clusters. Of the sample of ULXs that \cite{poutanen_2013} analyzed, four sources had offsets from clusters that were lower than the \textit{Chandra} astrometric accuracy. Of these, two were associated with the nuclei of the merging galaxies. The remaining two shared similar luminosities to the rest of the sample, and it's likely that they are from the same offset population, except offset towards our line of sight. This suggests that there are currently no observed accreting subsystems in the Antennae Galaxies, allowing us to place constraints on the black hole subsystem populations of YMCs. As shown in Figure \ref{fig:luminosityConditions}, if these systems harbor accreting subsystems, then they must have luminosities less than $\approx 5\times10^{37}\,{\rm ergs\,s^{-1}}$, which is  the approximate limiting luminosity of the Antennae Galaxies. Assuming a fixed velocity ratio of $\sigma/c_\infty = 4$ for a subsystem with a luminosity of this value, we can plot $M_{\rm BH}$ as a function of $N$ for different ambient conditions $\rho_\infty$ and $c_\infty$. We depict this in Figure \ref{fig:cuspConstraint}. Here, it is suggested that if $\rho_\infty = 10^{-22}\,{\rm g\,cm^{-3}}$ and $10\,{\rm km\,s^{-1}}<c_\infty<20\,{\rm km\,s^{-1}}$, as is expected in the Antennae Galaxies, then Antennae YMCs can hold no more than $\approx 5-20$, $\approx 10\,M_\odot$ black holes. In the case that the ambient density of the Antennae galaxies is $\sim 10^{-24}\,{\rm g\,cm^{-3}}$, which is typical of the Milky Way ISM, then the subsystem could be composed of no more than $\sim 60-200$, $10\,M_\odot$ black holes. We emphasize that at these values of $\sigma/c_\infty$ and $c_\infty$, the cluster mass is on the order of $10^5\,M_\odot$, indicating that our constraint would be even more restrictive for clusters masses of $\sim 10^6\,M_\odot$ which are also common in the Antennae Galaxies.  These are significant constraints which are in conflict with the predictions of dynamical simulations, which suggest that a typical subsystem population for similar clusters should be on the order of several hundred to thousands of black holes \citep[e.g.,][]{merritt_2004, mackey_2007, mackey_2008, SippelHurley2013, breenheggie_2013a, breenheggie_2013b, Ziosi2014, morscher_2015, ArcaSedda2016, peuten_2016, wang_2016, Banerjee2017, Askar_2018, arcasedda_2018, Webb2018, kremer_2019c, Weatherford_2019}.

\subsection{Conclusions}
\label{subsec:conclusion}
The constraints imposed in Section \ref{subsec:antennae} have potentially significant implications for the presence of mass-segregated black hole populations in Antennae YMCs, as they predict that YMCs host smaller black hole populations than is conventionally expected. We stress that while our analysis focused on the Antennae Galaxies, a similar analysis can be done for any system containing a large population of intermediate age ($\gtrsim50\,{\rm Myrs}$) YMCs. This includes other starburst galaxies, such as the Cartwheel Galaxy which has many observed ULXs \citep{gao_2003,wolter_trinchieri_2004}, and nearby gas-poor systems such as the Large Magellanic Cloud which is also rich in YMCs \citep{mackey_gilmore_2003}. Cluster simulations with masses and ages similar to those currently found in the Antennae Galaxies predict that typical subsystem populations are on the order of several hundreds to thousands of black holes. The discrepancy between this prediction and our results must either be: 1. due to an overestimation of black hole retention in young, solar metallicity clusters, 2. due to our overestimation of the amount of gas supplied to these clusters, or 3. due to a bandpass mismatch in existing ULX observations.

For completeness, we note that previous works have suggested that YMCs are generally devoid of gas \citep{Bastian2015,CabreraZiri2015,Longmore2015}. However, we emphasize that these works searched for molecular, star-forming gas and constrained the total gas mass to be $<9\%$ of the total cluster mass, where in our clusters the total gas mass is well within this constraint and is also expected to be ionized. Additionally, the study of YMC gas reservoirs is typically at YMC ages less than $50\,{\rm Myr}$, while we are interested in the gas reservoirs of intermediate-age YMCs (see Figure \ref{fig:cartoon}). 

One key uncertainty governing black hole retention in YMCs is the nature of black hole natal kicks. Current $N$-body cluster models typically assume that black holes are formed with mass fallback and calculate black hole natal kicks by sampling from the same kick distribution expected for core-collapse neutron stars \citep{hobbs_2005} but with the black hole kicks reduced in magnitude according to the fractional mass of fallback material \citep[see, e.g.,][for further details]{Fryer2012,morscher_2015}. However, the physics of black hole natal kicks, and in particular, the dependence upon metallicity is highly uncertain, and to date, poorly constrained observationally. If black holes receive stronger natal kicks at birth than is typically assumed in $N$-body cluster models, the black hole populations could be depleted from their host clusters at early times. Indeed, the lack of observed accreting subsystems in YMCs in the Antennae Galaxies may in fact place constraints upon the black hole formation process, including natal kicks, in these environments.

If on the other hand the reason for the discrepancy between our results and those of dynamical simulations instead lies with the details of our gas accumulation, the most likely scenario is that we underestimated the feedback processes that will impede accretion. In general, the feedback processes that could be included are: stellar feedback (from radiation and winds), black hole feedback (from radiation pressure and drag forces), and supernova feedback. We have chosen to focus on intermediate age YMCs \citep[$\gtrsim50\,{\rm Myr}$, $\approx 8\times10^3$ of which have been observed in the Antennae galaxies;][]{Whitmore_2010,fall_2005}, where the average time between supernovae is greater than the accretion timescale, suggesting that our results should be unaffected by supernova feedback. If there is radiation pressure from Eddington limited accreting black holes, this will induce a duty cycle in the luminous subsystem, but all subsystems will still be periodically visible. Realistically, black holes will move supersonically with respect to the accumulated gas and deposit energy via drag forces, but previous simulations of Bondi-Hoyle accretion onto binary systems suggest that these drag forces only marginally affect accretion rates \citep{antoni_2019}. While the luminosity field within any YMC is significant, our gas is optically thin and only weakly couples to it. Previous works have also shown that the ram pressure of stellar winds is much less than the accumulated gas pressure and is unlikely to impede the gas accumulation \citep{naiman_2011}. While we can contest any of these processes individually, it is possible that they can conspire to partially impede the cluster gas accumulation. This could lower our predicted cusp luminosities enough that our constraints are no longer in conflict with the predictions of dynamical simulations.

A final possibility is that we simply haven't observed the X-ray luminosity of these accreting subsystems at the right frequencies. The observations of ULXs that we have compared to were performed with \textit{Chandra}, which operates in the soft X-ray ($\approx0.5-10\,{\rm keV}$) regime. While black hole accretion disks are bright in the soft X-rays, the geometry of the accretion flow onto our black holes is quasi-spherical. This suggests that the the accretion flow may remain hot and radiatively-inefficient on small scales, which would lead to non-thermal hard X-ray emission at higher frequencies than \textit{Chandra} can detect. If this is the case, we should look to X-ray observatories that are sensitive to harder frequencies, such as \textit{NuSTAR}, to provide insights on these accreting black hole subsystems.

These uncertainties aside, we expect that given the large number ($\approx 8\times 10^3$) of observed YMCs in the Antennae galaxies that could viably host an accreting subsystem, the non-detection of such a luminous signal deserves attention. We argue that as such, luminous accreting subsystems should be considered as an observational diagnostic in future searches for black hole populations in dense star clusters. Further X-ray surveys of the Antennae galaxies and other young starbursts galaxies will continue to increase the constraints provided in Figure \ref{fig:cuspConstraint}. As these constraints grow stronger, it will be necessary to study the physics of accreting subsystems in more detail to better understand gas retention in YMCs. If our analysis holds up to future observations and studies that include feedback processes, then we may need to revisit our understanding of black hole retention in dense star clusters. 

\begin{acknowledgements}
We thank A. Antoni, D. Lin, J. Naiman and F. Rasio for guidance and encouragement. ER-R thanks the Heising-Simons Foundation, the Danish National Research Foundation (DNRF132) and NSF (AST-1911206 and AST-1852393) for support. KA is supported by the Danish National Research Foundation (DNRF132).  The authors thank the Niels Bohr Institute for its hospitality while part of this work was completed. The simulations presented in this work were performed using software produced by the Flash Center for Computational Science at the University of Chicago, which was visualized using code managed by the yt Project. \software{yt (Turk et al. 2011) \url{https://yt-project.org}, FLASH (Fryxell et al. 2000) \url{https://astro.uchicago.edu/research/flash.php}, Matplotlib (Hunter 2007) \url{https://matplotlib.org/3.3.3/index.html}}
 
\end{acknowledgements}

\bibliographystyle{aasjournal}

%\bibliography{references/references}

\end{document}